%% file: J-interp-approx.tex
\def\arxiv{1} %
\documentclass[twocolumn]{autart}
\usepackage{cite}[sort,compress]
\usepackage[pdftex]{graphicx}
\usepackage{array}
\usepackage{url}
\usepackage{arydshln}
\usepackage{float}
\usepackage{scrextend}
\usepackage{blindtext}
\usepackage{amsmath,amsfonts,amssymb}
\usepackage{tikz}
\usepackage{xspace}
\usepackage{xcolor}
\usepackage{mathrsfs}
\usepackage{mathalfa}
\usepackage{euscript}
\usepackage{csquotes}
\usepackage{datetime}
\usepackage{arydshln}
\usepackage{enumitem}
\usepackage{algorithm}
\usepackage{algpseudocode}
\algrenewcommand\algorithmicrequire{\textbf{Input}}

\algrenewcommand\algorithmicensure{\textbf{Output}}

\ifcsname arxiv\endcsname
\usepackage{xpatch}
\makeatletter %
\xpatchcmd{\runningauthor@fmt}{\global\edef}{\protected@xdef}{}{}
\xpatchcmd{\runningauthor@fmt}{\global\edef}{\protected@xdef}{}{}
\xpatchcmd{\author@fmt}{\edef}{\protected@edef}{}{}
\def\@xnamedef#1{\expandafter\protected@xdef\csname #1\endcsname}
\def\ead@au#1{\protected@edef\@ead@au{#1}}
\def\add@xtok#1#2{\begingroup
  \protected@xdef\@act{\global\noexpand#1{\the#1#2}}\@act
\endgroup}
\def\no@harm{}
\makeatother
\fi

\input{cc_intapp}

\begin{document}

\begin{frontmatter}

\ifx\arxiv\undefined\title{Direct data-driven interpolation and approximation of linear parameter-varying system trajectories} 
\else\title{Direct data-driven interpolation and approximation \mbox{of linear parameter-varying system trajectories}}\fi
 
\thanks{This work has been supported by the European Union within the framework of the National Laboratories for Autonomous Systems (RRF-2.3.1-21-2022-00002). Corresponding author: Chris Verhoek ({\tt c.verhoek@tue.nl}).}
 
\author[tue]{Chris Verhoek},    
\author[icrea,cimne]{Ivan Markovsky},           
\author[tue,sztaki]{Roland T{\'o}th}

\address[tue]{Control Systems Group, Eindhoven University of Technology, 5600MB Eindhoven, The Netherlands}
\address[icrea]{Catalan Institution for Research and Advanced Studies (ICREA), Pg. Lluis Companys 23, Barcelona, Spain}
\address[cimne]{Centre Internacional de M{\`e}todes Num{\`e}rics en Enginyeria (CIMNE), S/N 08034 Barcelona, Spain}
\address[sztaki]{Systems and Control Lab, HUN-REN Institute for Computer Science and Control, 1111 Budapest, Hungary}
          
\begin{keyword}
    Data-driven control theory, Behavioral approach, Linear parameter-varying systems, Missing data estimation.
\end{keyword}

\begin{abstract}
We consider the problem of estimating missing values in trajectories of \emph{linear parameter-varying} (LPV) systems. We solve this \emph{interpolation} problem for the class of shifted-affine LPV systems. Conditions for the existence and uniqueness of solutions are given and a direct data-driven algorithm for its computation is presented, i.e., the data-generating system is not given by a parametric model but is implicitly specified by data. We illustrate the applicability of the proposed solution on illustrative examples of a mass-spring-damper system with exogenous and endogenous parameter variation. 
\end{abstract}

\end{frontmatter}

\section{Introduction}\label{s:intro}

Increasing complexity of systems makes the development of models based on first principles difficult, costly, or sometimes even impossible. This also hinders the usage of traditional (model-based) methods for analysis and controller design. This problem, together with the increasing availability of data, has generated significant interest in direct data-driven analysis and control design. The behavioral approach~\cite{PoldermanWillems1997} provides a theoretical framework for this purpose, while also providing the possibility for giving rigorous guarantees on stability and performance. The behavioral approach views dynamical systems as a set of trajectories---signals that satisfy the dynamical laws of the system, which can be described by various types of representations (called \emph{realizations}).

A key result in the field of direct data-driven control is the Fundamental Lemma~\cite{WillemsRapisardaMarkovskyMoor2005}. It provides a data-based representation of the finite-horizon behavior of a \emph{Linear Time-Invariant} (LTI) system. This representation is constructed from a possibly single trajectory of the system, given that it is `rich' enough. Based on this result, a plethora of data-driven analysis and control methods have been developed for LTI systems~\cite{MarkovskyRapisarda2008, coulson2019deepc, romer2019one}. 

Particularly interesting applications of the data-based representation of LTI systems are data-driven interpolation and approximation of trajectories. Interpolation aims at finding missing values in a partially specified trajectory. It is a \emph{generalization} of simulation---the process of obtaining the output of the system from given input and initial conditions, which can also be viewed as finding missing values in a partially specified trajectory. 

Data-driven simulation is considered in the context of LTI control in~\cite{fujisaki2004system, ikeda2001model}, while in~\cite{markovsky2005data, MarkovskyRapisarda2008} the Fundamental Lemma is employed for data-driven simulation. Its generalization, the data-driven interpolation problem, is solved in~\cite{markovsky2022interpolation}.

The aforementioned results on data-driven simulation and interpolation are for LTI systems. However, with the  growing complexity of engineering problems, it is increasingly important to consider the nonlinear dynamical behavior of systems in practice. Existing works on direct data-driven simulation beyond the LTI case consider nonlinear systems with a specific structure, e.g., bilinear systems~\cite{markovsky2022data}, nonlinear ARX systems~\cite{mishra2021data}, or nonlinear systems that can be written in terms of a known basis function expansion~\cite{fazzi2024simulation}.

Alternatively, an efficient way of handling nonlinear behaviors is in the framework of \emph{Linear Parameter-Varying}~(LPV) systems. Behaviors of LPV systems are characterized by a linear dynamic relationship that depends on a \emph{measurable} signal~$p$, referred to as the \emph{scheduling signal}~\cite{verhoek2024encyclo}. The LPV framework serves as a bridge between the LTI framework and the realm of nonlinear systems. The data-driven simulation problem for the class of LPV systems that have a kernel representation with a \emph{Shifted-Affine} (SA) scheduling dependence has been solved in~\cite{verhoek2024flsimple}. The methods in~\cite{verhoek2023dpcjournal,verhoek2023dissipativity} have already applied this solution in data-driven predictive control and data-driven dissipativity analysis for LPV systems. However, the solution to the data-driven interpolation problem for LPV systems is still open. Its solution would open the door for solving problems on missing data recovery or trajectory planning. Contrary to the data-driven simulation problem, the interpolation problem may not have a unique solution, or have no solution at all. In this paper, we solve the LPV data-driven interpolation problem and derive necessary and sufficient conditions for existence and uniqueness of solution.

The remainder of the paper is structured as follows.
Section~\ref{s:preliminaries} provides the system class we consider and the necessary preliminaries.
The problem statement is given in Section~\ref{s:problem}, while its solution is derived in Section~\ref{s:int}.
We provide illustrative examples in Section~\ref{s:examples} and conclude the paper in Section~\ref{s:conclusion}.

\subsection*{Notation}

$\mb{R}$ is the set of real numbers, while the set of integers is given by $\mb{Z}$. 
For the (sub)spaces $\mb{A},\mb{B}$, %
the notation $\mb{B}^\mb{A}$ indicates the collection of all maps from $\mb{A}$ to $\mb{B}$. 
The $p$-norm of a vector $x\in\mathbb{R}^{n_\mathrm{x}}$ is denoted by $\lVert x\rVert_p$. For the two matrices $A\in\mb{R}^{n\times m}$ and $B\in\mb{R}^{p\times q}$, the Kronecker product is given as $A\kron B\in\mb{R}^{pm \times qn}$, while $\mathrm{blkdiag}$ is the block-diagonal composition for matrices, i.e., $\mathrm{blkdiag}(A,B) = \begin{bsmallmatrix} A & 0\\ 0 & B \end{bsmallmatrix}\in\mb{R}^{n+p\times m+q}$. The identity matrix of size $n\times n$ is denoted as $I_n$. Furthermore, $\col(x_1, \dots ,x_n)$ denotes the column vector $\begin{bsmallmatrix}x_1^\top & \cdots & x_n^\top \end{bsmallmatrix}^\top$. An index set $\left\{n,\, n+1,\, \dots,\, m\right\}$, with $n,m\in\mb{Z}$ and $n\leq m$, is denoted by $\mb{I}_n^m$.
Consider a signal $w: \mb{Z}\to\mb{R}^{\dnw}$. The value of a signal $w:\mb{Z}\to\mb{R}^{\dnw}$ at time step~$k$ is denoted as $w(k)\in\mb{R}^{\dnw}$ and its $i$\tss{th} element %
is given by $w_i(k)\in\mb{R}$. The forward and backward time-shift operators are denoted by $\q$ and $\q^{-1}$. We denote a time-interval between $t_1$ and $t_2$, $t_1\leq t_2$ by $[t_1,t_2]\subset\mb{Z}$. For the time interval $\mb{T}\subseteq\mb{Z}$, we write $w_\mb{T}$ as the truncation of $w$ to $\mb{T}$, e.g., for $\mb{T}:=[1,N]$ we have $w_{[1,N]}=(w(1),\dots,w(N))\in\left(\mb{R}^\dnw\right)^{[1,N]}$. {The notation $w_{[1,N_1]}\land v_{[1,N_2]}\in\left(\mb{R}^\dnw\right)^{[1,N_1+N_2]}$ indicates concatenation of  $w_{[1,N_1]}$ followed by $v_{[1,N_2]}\in\left(\mb{R}^\dnw\right)^{[1,N_2]}$}, while, with a slight abuse of notation, $\col(w, v)$ indicates the stacked signal $(\dots, \begin{bsmallmatrix} w(k-1) \\ v(k-1) \end{bsmallmatrix}, \begin{bsmallmatrix} w(k) \\ v(k) \end{bsmallmatrix}, \begin{bsmallmatrix} w(k+1) \\ v(k+1) \end{bsmallmatrix}, \dots)$. A sequence of the following form $(p(k)\kron w(k))_{k=1}^{N}$ is denoted by $w^\mt{p}_{[1,N]}$. For $w_{[1,N]}$, the associated Hankel matrix of depth~$L$ is given by
\[ 
    \mc{H}_L(w_{[1,N]})=\begin{bsmallmatrix}w(1) & w(2) & \dots & w({N-L+1})\\ w(2) & w(3) & \dots & w({N-L+2})\\ \vdots & \vdots & \ddots & \vdots \\ w({L}) & w({L+1}) & \dots & w({N})
\end{bsmallmatrix},
\]
while the block-diagonal Kronecker operator `$\bkron$' is denoted as ${w}_{[1,N]}\bkron I_n =\mathrm{blkdiag}\big(w(1)\kron I_n, \dots, w(N)\kron I_n\big)$. Finally, throughout the paper, we distinguish signals from {recorded} data sets with a breve accent, e.g., $\breve{w}$.

\section{Preliminaries}\label{s:preliminaries}

We consider the class of discrete-time LPV systems that can be represented by a kernel representation with \emph{shifted-affine} (SA) scheduling dependence:
\begin{subequations}\label{eq:lpvsys}
\begin{equation}
    {\sum_{i=0}^{\dnr}}r_i(\q^i p)\q^i w = 0,
\end{equation}
with manifest signals\footnote{\label{footnote:manifest}{The manifest signals are the signals of the system that interact with the environment, e.g., inputs and outputs.}} $w\in\left(\mb{R}^{\dnw}\right)^\mb{Z}$, scheduling signals $p\in\mb{P}^\mb{Z}$ with $\mb{P}\subseteq\mb{R}^\dnp$ the scheduling set, and scheduling dependent coefficient functions with a shifted time-dependence and an affine functional dependence:
\begin{equation}\label{eq:sa}
    r_i(\q^i p) = r_{i,0}+{\sum_{j=1}^{\dnp}} r_{i,j}\q^i p_j.
\end{equation}
\end{subequations}
The representation~\eqref{eq:lpvsys} is a representation of a given behavior~$\mf{B}$ of an LPV-SA system if 
\[ \mf{B}=\setdefinition{(w,p)\in(\mb{R}^{\dnw}\times\mb{P})^\mb{Z}}{\text{\eqref{eq:lpvsys} holds}}. \]
When needed, we consider an input/output partitioning of the variables $w=\Pi\begin{bsmallmatrix}u \\ y\end{bsmallmatrix}$, where $\Pi$ is a permutation matrix, $u\in(\mb{R}^\dnu)^\mb{Z}$ is the input, and $y\in(\mb{R}^\dny)^\mb{Z}$ is the output. The number of \emph{maximally free} inputs~$\mBf$ is a property of the behavior~$\mf{B}$, and it is invariant of the choice of the partitioning. Two other invariant properties of~$\mf{B}$ are the \emph{lag}~$\LBf$ and the \emph{order}~$\nBf$. The lag is the minimum number of lags, i.e.,~$\dnr$ in~\eqref{eq:lpvsys}, such that~\eqref{eq:lpvsys} can still characterize~$\mf{B}$. The order~$\nBf$ coincides with the minimum number of states that are required for a LPV-SS representation with static dependence that has a manifest behavior that coincides with~$\mf{B}$. We refer to LPV systems for which the behavior is characterized by a kernel representation of the form~\eqref{eq:lpvsys} as LPV-SA systems, and denote the class of LPV-SA systems with $\dnp$ scheduling signals and $\dnw$ manifest variables by $\syssa$.

We will consider some specific subsets of $\mf{B}$ throughout the paper:
\begin{subequations}
\begin{align}
    \mf{B}_\mb{P} & =  \{ p\in\mb{P}^\mb{Z} \mid \exists w\in\left(\mb{R}^{\dnw}\right)^\mb{Z} \text{ s.t. }(w,p)\in\mf{B}\}, \label{eq:BP} \\
    \mf{B}_p & = \{ w\in(\mb{R}^{\dnw})^\mb{Z} \mid (w,p)\in\mf{B}\}\label{eq:Bp},
\end{align}
providing the set of admissible scheduling trajectories of~$\mf{B}$ and the set of~$w$ trajectories that are compatible with a given  scheduling trajectory~$p\in\mf{B}_\mb{P}$, respectively. We are also interested in the trajectories in~$\mf{B}$ that are restricted to the time interval $[1,L]\subset\mb{Z}$, $1\leq L$:
\begin{multline}
    \Bfint{}{[1,L]}  = \big\{(w,p)_{[1,L]}\in(\mb{R}^{\dnw}\times\mb{P})^{[1,L]} \mid \exists \,(\omega,\rho)\in\mf{B} \\ \text{s.t. } (w(k),p(k))=(\omega(k),\rho(k)) \text{ for } 1\le k\le L\big\}.
\end{multline}
\end{subequations}
As proven in~\cite{verhoek2024flsimple}, the initial condition of a trajectory $(w_\mr{r}, p_\mr{r})\in\Bfint{}{[1,T_\mr{r}]}$ can be uniquely characterized by an \emph{initial trajectory} $(w_\mr{i},p_\mr{i})\in\Bfint{}{[1,T_\mr{i}]}$ that has a length $T_\mr{i}\geq\LBf$. This is summarized in the following lemma, adopted from~\cite[Lem.~2]{verhoek2024flsimple}.
\begin{lemma}\label{lem:state}
    Consider an LPV-SA system $\Sigma\in\syssa$ with behavior~$\mf{B}$. Given $(w_\mr{i},p_\mr{i})\in(\mb{R}^{\dnw}\times\mb{P})^{[1,T_\mr{i}]}$. If $T_\mr{i}\geq\LBf$ and \(
        (w_\mr{i},p_\mr{i})\land(w_\mr{r}, p_\mr{r})\in\Bfint{}{[1,T_\mr{i}+T_\mr{r}]},
    \)
    then the initial condition of $(w_\mr{r}, p_\mr{r})$ is uniquely expressed in terms of $(w_\mr{i},p_\mr{i})$, and equivalently, for a minimal LPV-SS representation of $\Sigma$ there exists a unique $\mt{x}\in\mb{R}^{\nBf}$ that serves as the initial condition of $(w_\mr{r}, p_\mr{r})$.
\end{lemma}
We are interested in interpolation of a $\Sigma\in\syssa$ directly from data, for which we require a data-driven representation of $\Sigma$. Hence, suppose we measure data from~$\Sigma$, collected in $\dataset = (\breve{w}_{[1,\Nd]}, \breve{p}_{[1,\Nd]})\in\Bfint{}{[1,\Nd]}$. 
Let $\breve{w}^{\breve{\mt{p}}}_{[1,\Nd]} = (\breve p(k)\kron\breve w(k))_{k=1}^{\Nd}$. Then, we have the following essential result from~\cite[Thm.~1]{verhoek2024flsimple}:
\begin{proposition}\label{prop:fl}
    Given a data-set $\dataset\in\Bfint{}{[1,\Nd]}$ from an LPV-SA system $\Sigma\in\Sigma_{\dnp,\dnw}$. Let $L\ge\LBf$. Then, {the following statements are equivalent:}
    \begin{enumerate}
    	\item For any $({w}_{[1,L]},{p}_{[1,L]})\in\Bfint{}{[1,L]}$, there exists a vector $g\in\mb{R}^{\Nd-L+1}$ that satisfies
	    \begin{equation}\label{eq:fundamentallemmafull}
	        \begin{bmatrix} \mc{H}_L(\breve w_{[1,\Nd]}) \\ \mc{H}_L(\breve{w}^{\breve{\mt{p}}}_{[1,\Nd]})-\mc{P}^\dnw \mc{H}_L(\breve w_{[1,\Nd]})\end{bmatrix}g=\begin{bmatrix} \mr{vec}(w_{[1,L]}) \\ 0 \end{bmatrix},
	    \end{equation}
	    with $\mc{P}^{\dnw}= p_{[1,L]}\bkron I_{\dnw}$.
	    \item {For every ${p}_{[1,L]})\in\Bfint{\mb{P}}{[1,L]}$,
		    \begin{equation}
		    	\mr{rank}\left(\mc{H}_L(\ddict{w}_{[1,\Nd]})\meu{N}_{p}\right) = \nBf + \mBf L, 
		    \end{equation}
		    where~$\meu{N}_{p}$ is a matrix whose columns form a basis for $\mr{kernel}(\mc{H}_L(\ddict{w}^{\breve{\mt{p}}}_{[1,\Nd]}) - \mc{P}^{\dnw}\mc{H}_L(\ddict{w}_{[1,\Nd]}))$.}
	    \item The following rank condition is satisfied:
		    \begin{equation}\label{eq:GPE}
		    \mr{rank}\left(\begin{bmatrix} \vphantom{\Big)} \mc{H}_L(\breve w_{[1,\Nd]}) \\ \mc{H}_L(\breve{w}^{\breve{\mt{p}}}_{[1,\Nd]}) \end{bmatrix}\right) = \nBf + (\mBf+\dnp\dnw)L.
		    \end{equation}
    \end{enumerate}
\end{proposition}
For a given~$p_{[1,L]}\in\Bfint{\mb{P}}{[1,L]}$, the left-hand side of~\eqref{eq:fundamentallemmafull} provides a \emph{data-driven representation} of the manifest behavior~$\Bfint{p}{[1,L]}$, under the condition that~\eqref{eq:GPE} holds, which is called the \emph{generalized persistence of excitation} (GPE) condition. We are now ready to state the problem formulation.

{\section{Problem formulation}}\label{s:problem}

In this paper, we solve the data-driven LPV interpolation problem: For a \emph{given} set of data points\footnote{These do not have to be given time-samples, they can also be given vector entries at certain times. See Example~\ref{ex:notation1} for clarification.}, the interpolation problem corresponds to finding the set of \emph{missing} data points from only measured data. The set of missing points composed \emph{together} with the given points, result in an admissible trajectory of the system. {Throughout the paper, we assume that the measured data is noise-free:}
\begin{assum}\label{ass:noise}
	{The measurements in $\dataset$ are noise-free.}
\end{assum}
Before formally defining the interpolation problem, we first introduce some necessary notation associated with it.

\subsection{{Given and missing values of $w_{[1,L]}$}}

Consider the trajectory $w_{[1,L]}\in(\mb{R}^{\dnw})^{[1,L]}$. Note that $\mr{vec}(w_{[1, L]})\in\mb{R}^{L\dnw}$. Now, consider the set of indices $\Idg\subseteq \mb{I}_1^{L\dnw}\subset\mb{N}$. Let the set of indices $\Idm\subseteq\mb{I}_1^{L\dnw}\subset\mb{N}$ be the set difference of $\mb{I}_1^{L\dnw}$ and $\Idg$, i.e. $\mb{I}_1^{L\dnw}\setminus\Idg$. The subscript $\mr{g}$ is for `given', while $\mr{m}$ is for `missing'. The notations $w_{\Idg}$ and $w_{\Idm}$ collect the given set of points and the missing (i.e., to-be-interpolated) points of the trajectory $w_{[1,L]}$, respectively. Let $\nIdg$ be the number of indices in $\Idg$. Similarly, we can apply these notations on Hankel matrices, e.g., $\mc{H}_L(\ddict{w}_{[1,\Nd]})|_{\Idg}\in\mb{R}^{\nIdg\times\Nd-L+1}$ selects the rows corresponding to the indices in $\Idg$. We clarify the notation with an example.
\begin{example}\label{ex:notation1}
    Given the trajectory $w_{[1,3]}\in(\mb{R}^2)^{[1,3]}$:
    \[
        w_{[1,3]}=\left(\begin{bsmallmatrix} 1\\4 \end{bsmallmatrix}, \begin{bsmallmatrix} 2\\5 \end{bsmallmatrix}, \begin{bsmallmatrix} 3\\6 \end{bsmallmatrix} \right),
    \]
    and let $\Idg = \{2,5,6\}$, i.e., $\nIdg=3$ and $\Idm = \{1,3,4\}$. Then, we have that $w_{\Idg}$ is given as
    \[ 
        w_{\Idg}= (4,3,6), 
    \]
    and $w_{\Idm}$ is given by
    \[ w_{\Idm}= (1,2,5). \]
    Now consider a trajectory from a data-dictionary:
    \[ \ddict{w}_{[1,5]}=\left(\begin{bsmallmatrix} 1\\2 \end{bsmallmatrix}, \begin{bsmallmatrix} 3\\4 \end{bsmallmatrix}, \begin{bsmallmatrix} 5\\6 \end{bsmallmatrix}, \begin{bsmallmatrix} 7\\8 \end{bsmallmatrix}, \begin{bsmallmatrix} 9\\0 \end{bsmallmatrix}\right), \]
    which, for $L=3$, gives
    \[ \mc{H}_3(\ddict{w}_{[1,5]}) = \begin{bsmallmatrix}
        1 & 3 & 5 \\ 
        2 & 4 & 6 \\
        3 & 5 & 7 \\
		4 & 6 & 8 \\
		5 & 7 & 9 \\
		6 & 8 & 0
    \end{bsmallmatrix}. \]
    Then, the matrix $\mc{H}_3(\ddict{w}_{[1,5]})|_{\Idg}$ is given by
    \[ \mc{H}_3(\ddict{w}_{[1,5]})|_{\Idg} = \begin{bsmallmatrix}
        2 & 4 & 6 \\
		5 & 7 & 9 \\
		6 & 8 & 0
    \end{bsmallmatrix}. \]
\end{example}
Finally, we introduce $\Idgp$, which is the set of indices that selects $w_{\Idgp}^\mt{p}$ from $w_{[1,L]}^\mt{p}$. With a slight misuse of notation, this selection operator can be understood as: $w_{\Idgp}^\mt{p} = \left(w_{\Idg}\right)^\mt{p}$. The values for $\Idgp$ can be found using the following formula:
\begin{multline}\label{eq:Idgp}
    \Idgp = \\ \mr{sort}\Big(\Big\{\!\big\{i+\dnw\big(r+(\lceil\tfrac{i}{\dnw}\rceil\!-\!1)(\dnp\!-\!1)-1\big)\big\}_{r=1}^{\dnp}\Big\}_{i=\Idg}\Big),
\end{multline}
with `$\mr{sort}$' the operator that sorts a set of integers in a  monotonically increasing order.
Similarly, we define $\Idmp$. We will also clarify this notation by means of an example:
\begin{example}
    Consider the $w$ trajectory as in Example~\ref{ex:notation1} and the scheduling trajectory
    \[
        p_{[1,3]} := \left( \begin{bsmallmatrix} 7\\10 \end{bsmallmatrix}, \begin{bsmallmatrix} 8\\11 \end{bsmallmatrix}, \begin{bsmallmatrix} 9\\12 \end{bsmallmatrix} \right).
    \]
    Hence, we have that
    \[
        w^\mt{p}_{[1,3]}=(p_k\kron w_k)_{k=1}^3 = \left(\begin{bsmallmatrix} 7\\28\\10\\40 \end{bsmallmatrix}, \begin{bsmallmatrix} 16\\40\\22\\55 \end{bsmallmatrix}, \begin{bsmallmatrix} 27\\54\\36\\72 \end{bsmallmatrix} \right).
    \]
    Using~\eqref{eq:Idgp}, we get that $\Idgp = \{2, 4, 9, 10, 11, 12\}$, 
    and thus
    \[
        w_{\Idgp}^\mt{p} = (28, 40, 27, 54, 36, 72).
    \]
\end{example}
Again, we can apply this notation on Hankel matrices, e.g., $\mc{H}_L(\ddict{w}^{\ddict{\mt{p}}}_{[1,\Nd]})|_{\Idgp}$. 

\subsection{{Problem statement}}
Now that we have established the notation, we can pose the LPV data-driven interpolation problem, which can be understood as the LPV generalization of the LTI data-driven interpolation problem,~cf.~\cite{markovsky2022data}. 
\begin{problem}\label{prob:interpolation}
    Given the data-dictionary $\dataset\in\Bfint{}{[1,\Nd]}$, a scheduling signal $\tilde{p}_{[1,L]}\in\Bfint{\mb{P}}{[1,L]}$, and a partially specified trajectory $w_{[1,L]}$ with given elements $\Idg$. Find a trajectory $\tilde{w}_{[1,L]}\in\Bfint{\tilde{p}}{[1,L]}$, such that $\tilde{w}_{\Idg} = w_{\Idg}$ and thus $(\tilde{w}_{[1,L]},\tilde{p}_{[1,L]})\in\Bfint{}{[1,L]}$. 
\end{problem}
We solve Problem~\ref{prob:interpolation} and provide example how the solution can be used for analysis and control of LPV-SA systems.

\section{LPV data-driven interpolation}\label{s:int}

From Problem~\ref{prob:interpolation}, we can derive three subquestions that we need to answer in order to obtain a solution to the LPV data-driven interpolation problem:
\begin{enumerate}[nosep]
    \item When is $\dataset$ sufficiently rich to characterize any arbitrary interpolant $\tilde{w}_{[1,L]}\in\Bfint{\tilde{p}}{[1,L]}$?
    \item When is $w_{\Idg}$ consistent with the data-generating system for a given scheduling signal $\tilde{p}_{[1, L]}$?
    \item When is the interpolant $\tilde{w}_{[1,L]}$ unique?
\end{enumerate}
The first question is answered by the rank condition in Proposition~\ref{prop:fl}, i.e.,~\eqref{eq:GPE}. Hence, the first condition required to hold for the solution of Problem~\ref{prob:interpolation} is:
\begin{condition}\label{con:int:repr}
    The data in the given data-dictionary $\dataset$ satisfy the GPE condition~\eqref{eq:GPE}.
\end{condition}
{This condition implies that the data is able to represent \emph{any} interpolant.}

The second question is answered by verifying whether there exists a trajectory $\tilde{w}_{[1,L]}\in\Bfint{\tilde{p}}{[1,L]}$ for which $w_{\Idg}=\tilde{w}_{\Idg}$. This can be verified by using the data-driven representation~\eqref{eq:fundamentallemmafull} and requiring that the vector 
\[ \begin{bmatrix}\mr{vec}(w_{\Idg})^\top & 0_{1\times L\dnp\dnw}\end{bmatrix}^\top\]
is in the column space of
\[ \begin{bmatrix} \mc{H}_L(\ddict{w}_{[1,\Nd]})|_{\Idg}  \\ \mc{H}_L(\ddict{w}_{[1,\Nd]}^{\ddict{\mt{p}}}) - \tilde{\mc{P}}\mc{H}_L(\ddict{w}_{[1,\Nd]}) \end{bmatrix}, \]
where $\tilde{\mc{P}} = \tilde{p}_{[1,L]}\bkron I_\dnw$. {Note that it is not sufficient to check whether only $w^{\tilde{\mt{p}}}_{\Idgp} = \tilde{w}^{\tilde{\mt{p}}}_{\Idgp}$ holds. In other words, it is \emph{not} sufficient to check whether $\begin{bsmallmatrix}\mr{vec}(w_{\Idg})^\top & 0_{1\times \dnp\nIdg}\end{bsmallmatrix}^\top$ is in the column space of 
\[ \begin{bmatrix} \mc{H}_L(\ddict{w}_{[1,\Nd]})|_{\Idg}  \\ \big(\mc{H}_L(\ddict{w}_{[1,\Nd]}^{\ddict{\mt{p}}}) - \tilde{\mc{P}}\mc{H}_L(\ddict{w}_{[1,\Nd]})\big)_{\Idgp} \end{bmatrix}. \]
This particular condition requires that there exist \emph{some} $\rho_{[1,L]}\in\Bfint{\mb{P}}{[1,L]}$ for which there exists an interpolant $\omega_{[1,L]}\in\Bfint{\!\rho}{[1,L]}$ that interpolates $w_{\Idg}$.} Therefore, to verify whether there exists an interpolant of $w_{\Idg}$ corresponding to the given scheduling signal $\tilde{p}_{[1,L]}$, the following condition must be satisfied.
\begin{condition}\label{con:int:exist}
    For the given $\dataset$ and $\tilde{p}_{[1,L]}\in\Bfint{\mb{P}}{[1,L]}$, the following relation holds
    \begin{multline*}
        \mr{rank}\left(\begin{bmatrix} \mc{H}_L(\ddict{w}_{[1,\Nd]})|_{\Idg}  \\ \mc{H}_L(\ddict{w}_{[1,\Nd]}^{\ddict{\mt{p}}}) - \tilde{\mc{P}}\mc{H}_L(\ddict{w}_{[1,\Nd]}) \end{bmatrix}\right) = \\ 
        \mr{rank}\left(\begin{bmatrix} \mc{H}_L(\ddict{w}_{[1,\Nd]})|_{\Idg} & \mr{vec}(w_{\Idg}) \\ \mc{H}_L(\ddict{w}_{[1,\Nd]}^{\ddict{\mt{p}}}) - \tilde{\mc{P}}\mc{H}_L(\ddict{w}_{[1,\Nd]}) & 0_{L\dnp\dnw \times 1} \end{bmatrix}\right).
    \end{multline*}
\end{condition}

{This second condition provides an existence condition for Problem~\ref{prob:interpolation}, and verifies whether the given trajectory samples are compatible with the projected behavior~$\Bfint{\tilde{p}}{[1,L]}$.}

The answer to the third subquestion lies in the amount of information that is in the rows of the Hankel matrices associated with~$\Idg$ and hence directly corresponds to the \emph{number} of given points $\nIdg$. A natural way to formulate a uniqueness condition for $\tilde{w}_{[1,L]}$ would be to check that the image of the Hankel matrices with the rows associated with $\Idm$ and $\Idmp$ removed still span the behavior of horizon~$\nIdg$. In other words,~\eqref{eq:GPE} should hold for the concatenated rows of the Hankel matrices corresponding to $\Idg$ and $\Idgp$. This condition, however, is too restrictive, because it enforces uniqueness of the interpolant for \emph{any} arbitrary scheduling signal $p_{[1,L]}\in\Bfint{\mb{P}}{[1,L]}$. Instead, we use~\cite[Thm.~1]{verhoek2024flsimple} to arrive at a uniqueness condition for the interpolant~$\tilde{w}_{[1,L]}$ that is associated with only the scheduling signal $\tilde{p}_{[1,L]}$. {Denote $N_{\tilde p}:= \mc{H}_L(\ddict{w}^{\breve{\mt{p}}}_{[1,\Nd]}) - \tilde{\mc{P}}\mc{H}_L(\ddict{w}_{[1,\Nd]})$ and define~$\meu{N}_{\tilde p}$ as a matrix whose columns form a basis for $\mr{kernel}(N_{\tilde p})$.} Then, we have the following condition for uniqueness:
\begin{condition}\label{con:int:unique}
    For the given data-dictionary $\dataset$ and~$\tilde{p}_{[1,L]}\in\Bfint{\mb{P}}{[1,L]}$, the following holds:
    \begin{multline}\label{eq:int:unique}
        \mr{rank}\left(\mc{H}_L(\ddict{w}_{[1,\Nd]})|_{\Idg}\meu{N}_{\tilde p}\right) = \\ \mr{rank}\left(\mc{H}_L(\ddict{w}_{[1,\Nd]})\meu{N}_{\tilde p}\right) = \nBf + \mBf L.
    \end{multline}
\end{condition}
{In short, this uniqueness condition requires that the rows of the Hankel matrix selected by $\Idg$ span the full behavior.}
If Conditions~\ref{con:int:repr},~\ref{con:int:exist}, and~\ref{con:int:unique} are satisfied, then any solution~$g$ of
\begin{equation}\label{eq:interp-sol-g}
    \begin{bmatrix} \mc{H}_L(\ddict{w}_{[1,\Nd]})|_{\Idg}  \\ \mc{H}_L(\ddict{w}_{[1,\Nd]}^{\ddict{\mt{p}}}) - \tilde{\mc{P}}\mc{H}_L(\ddict{w}_{[1,\Nd]}) \end{bmatrix}g = \begin{bmatrix} \mr{vec}({w}_{\Idg}) \\ 0_{\dnp\dnw\times1} \end{bmatrix}
\end{equation}
yields a unique interpolant $\tilde{w}_{[1,L]} = \mc{H}_L(\ddict{w}_{[1,\Nd]})g$ of $w_{\Idg}$, where $(\tilde{w}_{[1,L]},\tilde{p}_{[1,L]})\in\Bfint{}{[1,L]}$. The LPV data-driven interpolation procedure is summarized in Algorithm~\ref{alg:interp}.
\begin{algorithm}[t]\caption{LPV data-driven interpolation}\label{alg:interp}
    \begin{algorithmic}[1]
    \Require $\dataset$, $\tilde{p}_{[1,L]}$ and $w_{\Idg}$.
        \State Solve \eqref{eq:interp-sol-g} for $g$.
        \State Compute $\mr{vec}(\tilde{w}_{[1,L]}) = \mc{H}_L(\ddict{w}_{[1,\Nd]})g$.
    \Ensure $\tilde{w}_{[1,L]}$.
    \end{algorithmic} 
\end{algorithm}
The following proposition summarizes the conditions for existence and uniqueness of the interpolant $\tilde{w}_{[1,L]}$:
\begin{proposition}\label{prop:existenceuniqueness}
    Given the data-dictionary $\dataset\in\Bfint{}{[1,\Nd]}$, a scheduling signal $\tilde{p}_{[1,L]}\in\Bfint{\mb{P}}{[1,L]}$, and a partially specified $w_{\Idg}$. The interpolant $\tilde{w}_{[1,L]}$ of $w_{\Idg}$ obtained from Algorithm~\ref{alg:interp}:
    \begin{itemize}
        \item Exists and is a trajectory of $\Bfint{\tilde{p}}{[1,L]}$ if and only if Conditions~\ref{con:int:repr} and~\ref{con:int:exist} are satisfied, 
        \item Is unique and a trajectory of $\Bfint{\tilde{p}}{[1,L]}$ if and only if Conditions~\ref{con:int:repr},~\ref{con:int:exist}, and~\ref{con:int:unique} are satisfied.
    \end{itemize}
\end{proposition}
\begin{proof}
	{For existence, we need to show that (\emph{i})~$\tilde{w}_{[1,L]}\in\Bfint{\tilde{p}}{[1,L]}$, and (\emph{ii})~$w_{\Idg} = \tilde{w}_{\Idg}$. First note that if Condition~\ref{con:int:repr} holds, by Proposition~\ref{prop:fl} we have that any $\tilde{w}_{[1,L]}\in\Bfint{\tilde{p}}{[1,L]}$ can be represented by the data through~\eqref{eq:fundamentallemmafull}. Second, if Condition~\ref{con:int:exist} holds, by the Rouch\'e-Capelli theorem, there exists a solution~$g$ to~\eqref{eq:interp-sol-g}, making Step~1 of Algorithm~\ref{alg:interp} feasible and ensuring~$w_{\Idg} = \tilde{w}_{\Idg}$. Finally, for any solution~$g$ to~\eqref{eq:interp-sol-g}, computing $\tilde{w}_{[1,L]}$ via Step~2 of Algorithm~\ref{alg:interp} is equivalent to jointly solving~\eqref{eq:interp-sol-g} and~\eqref{eq:fundamentallemmafull}, i.e., by Proposition~\ref{prop:fl} $\tilde{w}_{[1,L]}\in\Bfint{\tilde{p}}{[1,L]}$. Hence, under Conditions~\ref{con:int:repr} and~\ref{con:int:exist}, $\tilde{w}_{[1,L]}$ interpolates $w_{\Idg}$ and $\tilde{w}_{[1,L]}\in\Bfint{\tilde{p}}{[1,L]}$. Uniqueness of $\tilde{w}_{[1,L]}$ follows from the fact that, under Condition~\ref{con:int:unique}, the null spaces of $\mc{H}_L(\ddict{w}_{[1,\Nd]})|_{\Idg}\meu{N}_{\tilde p}$ and $\mc{H}_L(\ddict{w}_{[1,\Nd]})\meu{N}_{\tilde p}$ are equivalent. Hence, under Conditions~\ref{con:int:repr}--\ref{con:int:unique}, any solution~$g$ to~\eqref{eq:interp-sol-g} will yield the same, i.e., unique, interpolant~$\tilde{w}_{[1,L]}$.}
\end{proof}
Condition~\ref{con:int:unique} provides a minimum on the number of points in~$\Idg$. We will, among others, discuss this in the following remark.
\begin{remark}\label{rem:minsamp}
    To have a unique interpolant $\tilde{w}_{[1,L]}$ for~$w_{\Idg}$, we can see from Condition~\ref{con:int:unique} that the minimum number of given points in $\Idg$, i.e., $\nIdg$, is $\nBf + \mBf L$. This, however, does not directly guarantee uniqueness, as $\mc{H}_L(\ddict{w}_{[1,\Nd]})|_{\Idg}\in\R^{\nIdg\times\Nd-L+1}$ may have \emph{linearly dependent} rows, and hence Condition~\ref{con:int:unique} might fail to hold. Additionally, note that if~\eqref{eq:int:unique} holds for~$\meu{N}_{\tilde{p}}|_{\Idgp}$, uniqueness is not guaranteed either. This is because~$\mc{H}_L(\ddict{w}_{[1,\Nd]})|_{\Idg}\meu{N}_{\tilde p}|_{\Idgp}$ characterizes all trajectories $(\hat{w},\hat{p})_{[1,L]}\in\Bfint{}{[1,L]}$ for which $\hat{w}_{\Idg}\equiv w_{\Idg}$ and $\hat{w}^{\hat{p}}_{\Idgp}\equiv w_{\Idgp}^{\tilde{p}}$. There are likely many~$\hat{p}_{[1,L]}\in\Bfint{\mb{P}}{[1,L]}$ for which this holds, hence it is important to consider the full scheduling signal~$\tilde{p}_{[1,L]}$ in verifying Condition~\ref{con:int:unique}. Finally, we want to make a note on the case when Condition~\ref{con:int:unique} is not satisfied, while Conditions~\ref{con:int:repr} and~\ref{con:int:exist} are. Let ${g}$ be a particular solution of Algorithm~\ref{alg:interp} and let
    \begin{equation}\label{eq:mcG}
        \mc{G} :=\mr{kernel}\left(\begin{bmatrix} \mc{H}_L(\ddict{w}_{[1,\Nd]})|_{\Idg} \\ N_{\tilde p} \end{bmatrix}\right).
    \end{equation}
    Then, $\mr{vec}(\tilde{w}_{[1,L]}) = \mc{H}_L(\ddict{w}_{[1,\Nd]})({g}+\hat{g})$ is an interpolant of~$w_{\Idg}$ for all~$\hat{g}\in\mc{G}$. Hence,~$\mc{G}$ specifies an infinite amount of valid interpolants of~$w_{\Idg}$.
\end{remark}
\begin{remark}\label{rem:sim}
    The LPV data-driven \emph{simulation} problem solved in~\cite{verhoek2024flsimple} can be seen as an important, but special case of the interpolation problem. In the simulation problem, the given points in~$\Idg$ correspond to the initial trajectory ($w_\mr{i}$ in Lemma~\ref{lem:state}) and the input corresponding to $w_\mr{r}$. {In terms of~\eqref{eq:mcG}, the rows of the Hankel matrix corresponding to $w_\mr{i}$ restrict $\mc{G}$ to contain only interpolants that are valid continuations from $w_\mr{i}$, while the rows of the Hankel matrices that correspond to the input signal in~$w_\mr{r}$ restrict $\mc{G}$ to contain only the interpolants that are a response to this input signal. Under a~$\dataset$ that satisfies the GPE condition for $L=T_\mr{i}+T_\mr{r}$ with $T_\mr{i}\geq\LBf$, the intersection of these spaces of interpolants gives a single, unique solution to the simulation problem.}
\end{remark}
\begin{remark}\label{rem:knownnBlB}
	{Verifying Conditions~\ref{con:int:repr},~\ref{con:int:exist}, and~\ref{con:int:unique} requires knowledge of the values for~$\nBf$, $\LBf$, and~$\mBf$. While the input dimension~$\mBf$ is generally known,~$\nBf$ and~$\LBf$ are typically unknown in practice. However, by realizing that for a given $p\in\Bfint{\mb{P}}{[1,L]}$,  $\Bfint{p}{[1,L]}$ is a linear subspace whose dimension is an affine function of~$L$, the values for~$\nBf$ and~$\LBf$ can be obtained from data following \cite[III.A]{markovsky2022identifiability}.}
\end{remark}
The solution to the interpolation problem opens up interesting extensions and generalizations when we relax some of the conditions, as discussed in the following remarks.
\begin{remark}
    If Condition~\ref{con:int:exist} cannot be satisfied, we can formulate the interpolation problem as an \emph{approximation} problem, where we aim to find a~$\tilde{w}_{[1,L]}$ that is the optimal approximator of $w_{\Idg}$, e.g., in a weighted least-squares sense. See~\cite{markovsky2022data} for a formulation for the LTI case. Hence, the data-driven LPV \emph{approximation} problem is formulated as follows:
    \begin{subequations}\label{eq:approx}
    \begin{align}
        \min_{\tilde{w}_{[1,L]}} \quad & \left\|\tilde{w}_{\Idg}-w_{\Idg}\right\|, \\
        \mr{s.t.}\quad & \tilde{w}_{[1,L]} \in\Bfint{\tilde{p}}{[1,L]},
    \end{align}
    \end{subequations}
    where we can take any norm in the cost. For example, the weighted $2$-norm of the approximation error, i.e.,
    \begin{equation}\label{eq:weighted2norm}
        \left\|\tilde{w}_{\Idg}-w_{\Idg}\right\| := \sqrt{(*)^\top M(\tilde{w}_{\Idg}-w_{\Idg})},
    \end{equation}
    with $M=M^\top\in\mb{R}^{\nIdg\times\nIdg}$ a positive definite weighting matrix. Let the columns of $\mc{B}_{\tilde{p}}$ form a basis for $\mr{image}\left(\mc{H}_L(\ddict{w}_{[1,\Nd]})\meu{N}_{\tilde p}\right)$. Then, the solution of~\eqref{eq:approx} with~\eqref{eq:weighted2norm} can be written explicitly as:
    \[ \mr{vec}(\tilde{w}_{[1,L]}) = \mc{H}_L(\ddict{w}_{[1,\Nd]})\left(M^{\frac{1}{2}}\mc{B}_{\tilde{p}}|_{\Idg}\right)^\dagger M^{\frac{1}{2}} w_{\Idg}. \]
\end{remark}
\begin{remark}
    It is also possible to relax Condition~\ref{con:int:repr} in terms of considering noisy data-dictionaries, i.e., the case where {Assumption~\ref{ass:noise} does not hold and thus} $\dataset\notin\Bfint{}{[1,\Nd]}$. This is handled in~\cite{markovsky2022data} by employing an $\ell_1$-norm regularization on~$g$ in the first step of Algorithm~\ref{alg:interp}. This relaxation is generally referred to as the Lasso regression~\cite{tibshirani1996regression}. Similar modifications may also work for the LPV case. However, these are largely still heuristic and further work should be done to achieve systematic noise handling for LTI and LPV systems.
\end{remark}
\begin{remark}
    We want to emphasize that the solution for the interpolation problem is directly applicable for LPV systems that are an embedding of a nonlinear system. To achieve this, we require a so-called \emph{scheduling map} $\psi:\mb{W}\to\mb{P}$ that maps the nonlinearities of the nonlinear system to the measurable scheduling signal $p$, see~\cite{verhoek2024encyclo} for further details on the embedding procedure. Through this scheduling map, we can apply iterative sequential quadratic programming methods~\cite{hespe2021convergence}, or nonlinear optimization methods to solve Algorithm~\ref{alg:interp} for nonlinear systems. The details of these interesting approaches, however, will not be discussed further, as they are outside of the scope of this paper.
\end{remark}
\begin{remark}
    The interpolation problem does not impose \emph{which} points are given. This means that we can also pose it, for example, as an (open-loop) control problem, where~$w_{\Idg}$ contains an initial trajectory and desired way points that the output of the system must reach. Moreover, due to the elegant formulation of Algorithm~\ref{alg:interp}, we can easily add constraints to the problem and arrive at a fully data-driven trajectory planning algorithm with the LPV data-driven interpolation problem at its core. We will demonstrate the efficiency of the proposed solution in solving such a control problem in the next section.
\end{remark}

\section{Examples}\label{s:examples}
We give two short examples to demonstrate the effectiveness of our proposed solution to the LPV data-driven interpolation problem on a simple parameter-varying \emph{mass-spring-damper} (MSD) system, taken from~\cite{verhoek2024flsimple}. The code for the examples can be found at https://gitlab.com/releases-c-verhoek/dd-interp-approx.

\subsection{System description}\label{ss:ex:sys}
The MSD system has a stiffness parameter that varies along a measurable signal~$p$, characterized by the affine relation
\[ \kappa(p(k)) = \kappa_0+\kappa_1 p(k), \quad p(k)\in[-1,1]. \]
To come up with an approximative DT representation of the CT MSD dynamics, we discretize the equations of motion using Euler's method with sampling time $\tau$. With $w=\mr{col}(u,y)$, the resulting DT form of the parameter-varying MSD system is described by~\eqref{eq:lpvsys}, where
\begin{multline}
    R(p,\q) = \\ \left[\begin{array}{l;{2pt/2pt}r} -\tfrac{\tau^2}{m} \ & \ \q^2 + \tfrac{d\,\tau-2m}{m}\q + \tfrac{m+\kappa_0\,\tau^2-d\,\tau}{m} + \tfrac{\kappa_1\,\tau^2}{m}p \end{array}\right],
\end{multline}
with $m$ the mass and $d$ the damping coefficient. We take $m=d=1$, $\kappa_0=10$, $\kappa_1=9$, and $\tau = 0.1$ in this paper. Note that $\dnw=2$, $\dnp=1$, $\nBf=2$, $\LBf=2$, and $\mBf=1$. 

\subsection{Example 1: Interpolation}
In these examples, we consider trajectories of length $L=30$, meaning that we need at least $\Nd=121$ samples in our data-dictionary~\cite[Eq.~(31)]{verhoek2024flsimple}. To achieve \emph{data-driven} interpolation, we generate a data-dictionary by applying $u(k)\sim\mc{N}(0,1)$ and $p(k)\sim\mc{U}(-1,1)$ to the MSD system (not depicted). A posteriori verification gives that the resulting $\dataset$ satisfies the GPE condition~\eqref{eq:GPE}.

\begin{figure}
    \centering
    \includegraphics[width=\linewidth]{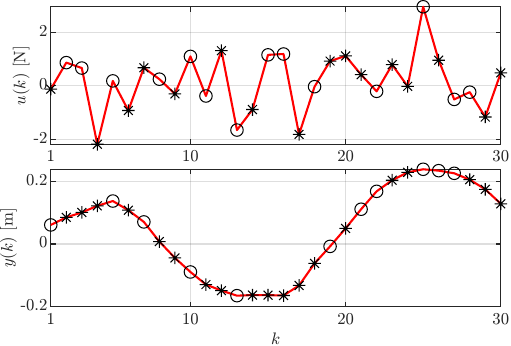}
    \caption{The true trajectory~$w_{[1,L]}$ of the MSD system (\legendline{red}), the given points~$w_{\Idg}$ (indicated by~$\ast$), and the missing points~$w_{\Idm}$ (indicated by~$\circ$). }\label{fig:int_a1}
\end{figure}

\begin{figure}
    \centering
    \includegraphics[width=\linewidth]{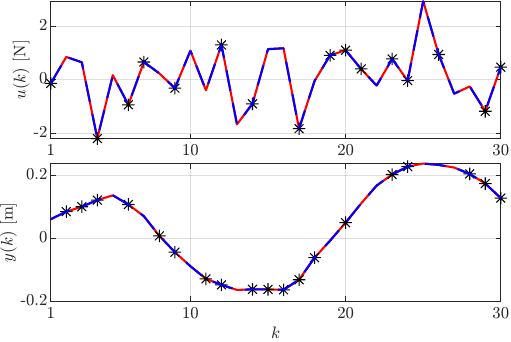}
    \caption{The true trajectory~$w_{[1,L]}$ of the MSD system~\mbox{(\legendline{red})}, the given points~$w_{\Idg}$ (indicated by~$\ast$), and the interpolated trajectory~$\tilde{w}_{[1,L]}$ (\legendlined{blue}) that is obtained from Algorithm~\ref{alg:interp}.}\label{fig:int_a2}
\end{figure}

From Remark~\ref{rem:minsamp}, we know that we need at least $\nBf + \mBf L=32$ points in $\Idg$ to achieve unique interpolation. However, in practice, it is often difficult to satisfy Condition~\ref{con:int:unique} for $\nIdg= \nBf + \mBf L$. We therefore choose a higher number of points ($\nIdg=35$) for interpolation. We randomly selected $35$ points from $\mb{I}_1^{L\dnw}$ to obtain $\Idg$. In Fig.~\ref{fig:int_a1}, we depicted the points in $\Idg$ and $\Idm$, together with the `true' trajectory. 
 With the given~$w_{\Idg}$ and the data-dictionary, simple rank-based computations verify that Conditions~\ref{con:int:repr}, \ref{con:int:exist}, and~\ref{con:int:unique} hold.
Solving the data-driven interpolation problem with Algorithm~\ref{alg:interp} will hence provide a unique trajectory~$\tilde{w}_{[1,L]}$ that interpolates~$w_{\Idg}$. The interpolated trajectory~$\tilde{w}_{[1,L]}$ is plotted in blue in Fig.~\ref{fig:int_a2}, together with~$w_{\Idg}$ and~$w_{[1,L]}$. 
The validity of Proposition~\ref{prop:existenceuniqueness} is verified in this example, as the interpolated trajectory is equivalent to the true trajectory, up to numerical precision. 

\begin{figure}
    \centering
    \includegraphics[width=\linewidth]{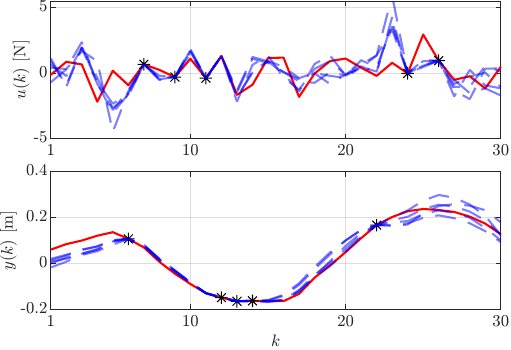}
    \caption{The true trajectory~$w_{[1,L]}$ of the MSD system~\mbox{(\legendline{red})} displayed with the given points~$w_{\Idg}$ (indicated by~$\ast$), where $\nIdg = 10$. Additionally, some valid interpolants of~$w_{\Idg}$ are shown in blue (\legendlined{blue}). For this case, Condition~\ref{con:int:unique} is not satisfied and there are multiple (in fact, an infinite number of) valid solutions for~$\tilde{w}_{[1,L]}$, characterized as in Remark~\ref{rem:minsamp}. }\label{fig:int_b}
\end{figure}

It is also worth to explore the scenario in which Condition~\ref{con:int:unique} is not met, e.g., when we take~$\nIdg = 10$ points. In this case, the interpolant is not unique and there exists an infinite amount of valid interpolants of~$w_{\Idg}$. We solve Algorithm~\ref{alg:interp} and compute the set of valid solutions~$g$ that provide interpolants of $w_{\Idg}$ (see also the discussion in Remark~\ref{rem:minsamp}). Fig.~\ref{fig:int_b} shows some valid solutions of the data-driven interpolation problem for the case when Condition~\ref{con:int:unique} is not satisfied.
Selecting the interpolant by taking linear combinations of vectors in $\mc{G}$, cf.~\eqref{eq:mcG}, allows for \emph{shaping} $\tilde{w}_{[1,L]}$, and hence opens the door for data-driven approximation methods, data-driven optimal control, etc. This is why the interpolation problem is such a fundamental and important problem. We will demonstrate a control application in the next example.

\subsection{Example 2: Control application}
So far, we considered the points in~$w_{\Idg}$ to be points of an existing trajectory for which we wanted to find the `remainder' of the trajectory. As highlighted in the previous paragraph, the interpolation problem also opens up possibilities for control applications. In such a situation, the points in~$w_{\Idg}$ are points that we want to \emph{reach}, and we want to find a control input such that we `interpolate' our `way-points' in an optimal sense. {Note that this results in a path planning and/or feedforward control problem. In this example, we want to design an input trajectory, i.e., a control input, for the system that interpolates a set of predefined desired points, which are collected in~$w_{\Idg}$.}

\subsubsection{Application for LPV systems} 
\begin{figure}
    \centering
    \includegraphics[width=\linewidth]{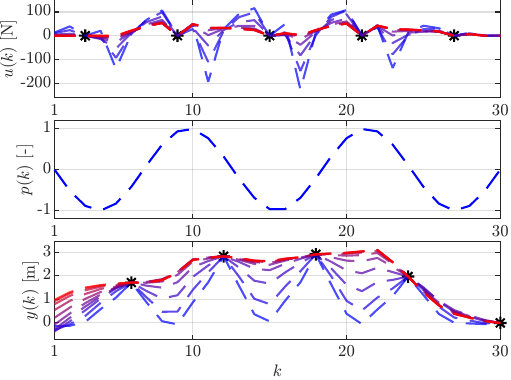}
    \caption{Solving the interpolation problem in a data-driven control context, where we solved \eqref{eq:intcon} for (exponentially) increasing $Q$, shown in a gradient from blue ($Q=10^{-4}$) to red ($Q=1$). The reference points for~$u$ and~$y$ are indicated with `$\ast$'.}\label{fig:int_con}
\end{figure}

Suppose that, under a sinusoidal scheduling signal, the output of the MSD needs to follow the parabolic curve $y_\mr{ref}(k) = \tfrac{-1}{70}k^2 + \tfrac{31}{70} k - \tfrac{3}{7}$ for the time-steps $k=6i$, $i\in\mb{N}$. Additionally, the input $u(k)$ must be zero at $k=6i-3$. We collect these points in $w_{\Idg}$. As an additional objective, the energy of $u$ must be minimized in overall. {Hence, the control objective is as follows:} for a given scheduling and reference points in $w_{\Idg}$, find the minimal-energy input signal that realizes the reference as an output response and satisfies $u(6i-3)\equiv 0$ for $i\in\mb{N}$. We can pose this as an optimal control problem where the reference points for~$u$ and~$y$ are collected in $w_{\Idg}$. We define a cost function:
\[ J(g) = g^\top \mc{H}_L(\ddict{w}_{[1,\Nd]})^\top \left(I_L\kron\begin{bsmallmatrix} Q & 0 \\ 0 & R \end{bsmallmatrix}\right) \mc{H}_L(\ddict{w}_{[1,\Nd]})\, g%
, \]
with $Q\in\mb{R}^{\dnu\times\dnu}$ and $R\in\mb{R}^{\dny\times\dny}$ as tuning matrices. %
Next, we solve the interpolation problem as an optimization problem, where the reference points in $w_{\Idg}$ are part of the constraints:
\begin{equation}\label{eq:intcon}
    \min_g \quad  J(g), \quad  \text{ subject to \eqref{eq:interp-sol-g}}.
\end{equation}
We show several solutions of the optimization in Fig.~\ref{fig:int_con} for $R=1$ and different values for $Q$.
The results in Fig.~\ref{fig:int_con} show that the solution to the LPV data-driven interpolation can indeed be used to design control inputs for LPV systems in a fully data-driven setting.

\subsubsection{Application for nonlinear systems}
\begin{figure}
    \centering
    \includegraphics[width=\linewidth]{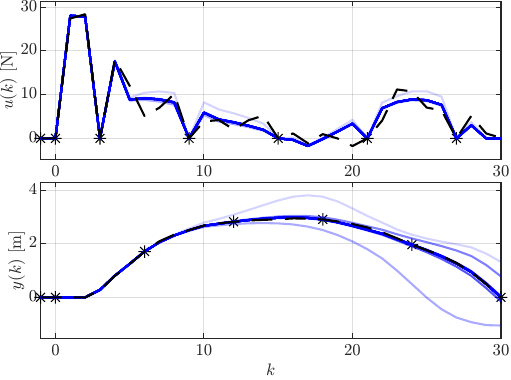}
    \caption{Solving the interpolation problem for an LPV embedding of a nonlinear system in the data-driven control context of Example~2. The solution obtained with {\tt fmincon} is shown in black (\legendlined{black}). The solution obtained using the SQP method is shown in blue~\mbox{(\legendline{blue})}. The (semi) transparent blue lines correspond to intermediate solutions of the iterative SQP method, where a higher opacity corresponds to a higher iteration. The reference points for~$u$ and~$y$ are indicated with `$\ast$'.}\label{fig:int_nl}
\end{figure}

As discussed in Section~\ref{s:intro}, LPV systems are often used in practice as surrogate descriptions of nonlinear systems. In such a case, the scheduling signal~$p$ captures the nonlinearities and time-variations of the underlying nonlinear system. Again, we consider the system described in Section~\ref{ss:ex:sys}, but in this case the scheduling signal is endogenous, i.e., $p(k)$ is functionally dependent on $w$:
\begin{equation}\label{eq:schedulingmap}
    p(k) = \cos(w_2(k))\tanh(w_2(k)).
\end{equation}
{We consider the same control problem and reference points~$w_{\Idg}$ as in the LPV case. We now, however, need to solve a \emph{nonlinear} control problem.} For this case, we choose $Q=R=1$ {and also added a zero initial trajectory (of length $\LBf$) to $w_{\Idg}$}. Now, $\tilde{\mc{P}}$ in~\eqref{eq:interp-sol-g} is nonlinearly dependent on $g$, i.e.,~\eqref{eq:intcon} is a nonlinear optimization problem. We solve the nonlinear optimization problem using two different methods. First, we use the interior point solver of {\tt fmincon} in \matlab with standard settings to solve~\eqref{eq:intcon}. Second, we use an iterative LPV method, reminiscent to \emph{sequential quadratic programming} (SQP), which is an approach similar to the methods presented in~\cite{hespe2021convergence} (see also~\cite{verhoek2024flsimple, verhoek2023dpcjournal} for similar applications). In the second approach, we take an estimate for~$w_{[1,L]}$ and propagate it through~\eqref{eq:schedulingmap} to obtain an estimate of~$p_{[1,L]}$. Then, we solve~\eqref{eq:intcon} for the estimated~$p_{[1,L]}$, i.e.,~\eqref{eq:intcon} is a convex quadratic program. Using the obtained solution~$w^{\mr{sol}}_{[1,L]}$ from~\eqref{eq:intcon}, we update the estimate of~$p_{[1,L]}$ and again solve~\eqref{eq:intcon} for the updated~$p_{[1,L]}$. We repeat this procedure until the solution converges. For the initial estimate of~$w_{[1,L]}$, we take a linear interpolation of~$w_{\Idg}$. We plotted the results of the two solution methods in Fig.~\ref{fig:int_nl}. {First note that, contrary to the results in Fig.~\ref{fig:int_con}, the solution trajectories all start from the same initial conditions, as highlighted in Remark~\ref{rem:sim}.} The solution obtained with {\tt fmincon} is shown in black. We have plotted the solution using the SQP-like method in blue. Additionally, we plotted several iterations of the SQP method in transparant blue, where a higher opacity corresponds to a higher iteration number. 
We terminated the iterative SQP method when $\|p_{[1,L]}-p^{\mr{prev}}_{[1,L]}\|<10^{-7}$, which occurred after 14~iterations. We computed the optimal~$J(g)$ for the obtained solutions, giving $2808$ for {\tt fmincon} and $2756$ for SQP. Computing $\|w_{\Idg}-w^{\mr{sol}}_{\Idg}\|$ gives $1.6\cdot10^{-4}$ for {\tt fmincon} and $8.6\cdot10^{-4}$ for SQP. This example shows that we can also effectively handle data-driven interpolation problems that involve LPV embeddings of nonlinear systems.

\remove{
\section{Other applications of the interpolation problem}
These are some ideas

\subsection{Scheduling (co)-interpolation}

problem: given $w$, find the (set of) scheduling signals $p$ such that $(w,p)\in\mf{B}$.

problem: given partial $(w,p)_{\Idg}$ (so $p$ is not given in full), now solve the interpolation problem. -> Becomes a bilinear problem, non-convex(?)

\subsection{Approximation: Inconsistent $w_{\Idg}$}

Condition~\ref{con:int:exist} is not satisfied.

-- trivial, get it for free

\subsection{Approximation: Noisy $\dataset$}

The assumption that $\dataset\in\Bfint{}{[1,\Nd]}$ is not satisfied.

assume though that the scheduling is noise-free?
}

\section{Conclusions}\label{s:conclusion}
In this paper, we solve a data-driven interpolation problem for LPV systems. Our proposed methodology is based on simple linear algebra operations, leading to general, practical and efficient algorithms that allow us to solve nontrivial problems, such as missing data recovery and trajectory planning for LPV systems. We demonstrated the power of the interpolation principle by illustrative examples, which also shows that we can use the developed methods for nonlinear systems. Future work involves extending the methods for noisy data and interpolation under inconsistent signal and scheduling trajectories.

\ifx\arxiv\undefined\bibliographystyle{ifacconf}\else\bibliographystyle{ieeetr}\fi
\bibliography{refs_intapp}

\end{document}

%% file: cc_intapp.tex
\DeclareRobustCommand{\legendline}[1]{\hspace{-0pt}\tikz[#1,line width=0.4mm,baseline=-0.5ex]{\draw (0,0) -- (.35,0);}\hspace{-0pt}}
\DeclareRobustCommand{\legendlined}[1]{\hspace{-0pt}\tikz[#1,line width=0.4mm,baseline=-0.5ex]{\draw[dashed] (0,0) -- (.3,0);}\hspace{-0pt}}

\definecolor{mblue}{rgb}{0,0.4470,0.7410}
\definecolor{morange}{rgb}{0.8500,0.3250,0.0980}
\definecolor{myellow}{rgb}{0.9290,0.6940,0.1250}
\definecolor{mpurple}{rgb}{0.4940,0.1840,0.5560}
\definecolor{mgreen}{rgb}{0.4660,0.6740,0.1880}
\definecolor{mcyan}{rgb}{0.3010,0.7450,0.9330}
\definecolor{mred}{rgb}{0.6350,0.0780,0.1840}
\definecolor{mgreenblue}{rgb}{0.0,1.0,0.5}
\definecolor{parulablue}{rgb}{0.2431,0.1490,0.6588}
\definecolor{parulalblue}{RGB}{39,151,235}
\definecolor{parulagreen}{RGB}{129,204,89}
\definecolor{parulayellow}{RGB}{249,251,21}
\definecolor{cblue}{rgb}{0,0.9,1}
\definecolor{corange}{rgb}{1,0.7,0}
\definecolor{mgray}{rgb}{0.8,0.8,0.8}

\ifx\replyletter\undefined%

\newenvironment{lemma}{\begin{lem}}{\end{lem}}

\newenvironment{example}{\begin{exmp}}{\end{exmp}}
\newenvironment{remark}{\begin{rem}}{\end{rem}}
\newenvironment{problem}{\begin{prob}}{\end{prob}}
\newenvironment{proposition}{\begin{prop}}{\end{prop}}
\newenvironment{proof}{\begin{pf}}{\hfill$\blacksquare$\end{pf}}
\newtheorem{condition}[thm]{Condition}
\fi

\newcommand{\remove}[1]{}

\newcommand{\setdefinition}[2]{\left\{\vphantom{#2}#1\right.\left|\,\vphantom{#1}#2\right\}}

\DeclareFontFamily{OT1}{pzc}{}
\DeclareFontShape{OT1}{pzc}{m}{it}{ <-> s*[1.1] pzcmi7t }{}
\DeclareMathAlphabet{\mathpzc}{OT1}{pzc}{m}{it}

\newenvironment{bsmallmatrix}%
    {\left[\begin{smallmatrix}}%
    {\end{smallmatrix}\right]}
    {\left(\begin{smallmatrix}}%
    {\end{smallmatrix}\right)}
    
\makeatletter
\DeclareRobustCommand\vdots{%
  \mathpalette\@vdots{}%
}
\newcommand*{\@vdots}[2]{%
  \sbox0{$#1\cdotp\cdotp\cdotp\m@th$}%
  \sbox2{$#1.\m@th$}%
  \vbox{%
    \dimen@=\wd0 %
    \advance\dimen@ -3\ht2 %
    \kern.5\dimen@
    \dimen@=\wd2 %
    \advance\dimen@ -\ht2 %
    \dimen2=\wd0 %
    \advance\dimen2 -\dimen@
    \vbox to \dimen2{%
      \offinterlineskip
      \copy2 \vfill\copy2 \vfill\copy2 %
    }%
  }%
}
\DeclareRobustCommand\ddots{%
  \mathinner{%
    \mathpalette\@ddots{}%
    \mkern\thinmuskip
  }%
}
\newcommand*{\@ddots}[2]{%
  \sbox0{$#1\cdotp\cdotp\cdotp\m@th$}%
  \sbox2{$#1.\m@th$}%
  \vbox{%
    \dimen@=\wd0 %
    \advance\dimen@ -3\ht2 %
    \kern.5\dimen@
    \dimen@=\wd2 %
    \advance\dimen@ -\ht2 %
    \dimen2=\wd0 %
    \advance\dimen2 -\dimen@
    \vbox to \dimen2{%
      \offinterlineskip
      \hbox{$#1\mathpunct{.}\m@th$}%
      \vfill
      \hbox{$#1\mathpunct{\kern\wd2}\mathpunct{.}\m@th$}%
      \vfill
      \hbox{$#1\mathpunct{\kern\wd2}\mathpunct{\kern\wd2}\mathpunct{.}\m@th$}%
    }%
  }%
}
\makeatother

\newcommand{\tss}[1]{\textsuperscript{#1}}

\newcommand{\matlab}{\textsc{Matlab}\xspace}

\newcommand{\mc}[1]{\mathcal{#1}}
\newcommand{\mr}[1]{\mathrm{#1}}
\newcommand{\mf}[1]{\mathfrak{#1}}
\newcommand{\mb}[1]{\mathbb{#1}}

\newcommand{\meu}[1]{\EuScript{#1}}

\newcommand{\mt}[1]{\mathtt{#1}}
\newcommand{\mbf}[1]{\mathbf{#1}}

\newcommand{\dny}{{n_\mr{y}}}
\newcommand{\dnu}{{n_\mr{u}}}
\newcommand{\dnp}{{n_\mr{p}}}
\newcommand{\dnw}{{n_\mr{w}}}

\newcommand{\dnr}{{n_\mr{r}}}

\newcommand{\R}{\mb{R}}

\newcommand{\Nd}{{N_\mr{d}}}
\newcommand{\dataset}{\mc{D}_\Nd}

\newcommand{\kron}{\otimes} %
\newcommand{\bkron}{\circledcirc}

\newcommand{\col}{\mr{col}}

\newcommand{\q}{\mr{q}}

\newcommand{\sysgen}{\Sigma}

\newcommand{\syssa}{\sysgen_{\dnp,\dnw}}

\newcommand{\Bfint}[2]{\left.\mf{B}_{#1}\right|_{#2}}

\newcommand{\nBf}{\mbf{n}(\mf{B})}
\newcommand{\LBf}{\mbf{L}(\mf{B})}

\newcommand{\mBf}{\mbf{m}(\mf{B})}

\newcommand{\Idg}{\mb{I}_\mr{g}}
\newcommand{\Idgp}{\mb{I}_\mr{g}^\mt{p}}
\newcommand{\Idm}{\mb{I}_\mr{m}}
\newcommand{\Idmp}{\mb{I}_\mr{m}^\mt{p}}
\newcommand{\nIdg}{K}

\newcommand{\ddict}[1]{{\breve{#1}}}

%% file: J-interp-approx.bbl
\begin{thebibliography}{10}

\bibitem{PoldermanWillems1997}
J.~W. Polderman and J.~C. Willems, {\em Introduction to Mathematical Systems
  Theory: A Behavioral Approach}, vol.~26.
\newblock Springer, 1997.

\bibitem{WillemsRapisardaMarkovskyMoor2005}
J.~C. Willems, P.~Rapisarda, I.~Markovsky, and B.~L.~M. De~Moor, ``A note on
  persistency of excitation,'' {\em Systems \& Control Letters}, vol.~54,
  no.~4, pp.~325--329, 2005.

\bibitem{MarkovskyRapisarda2008}
I.~Markovsky and P.~Rapisarda, ``Data-driven simulation and control,'' {\em
  International Journal of Control}, vol.~81, no.~12, pp.~1946--1959, 2008.

\bibitem{coulson2019deepc}
J.~Coulson, J.~Lygeros, and F.~D{\"o}rfler, ``Data-enabled predictive control:
  in the shallows of the {DeePC},'' in {\em Proc. of the 2019 European Control
  Conference}, pp.~307--312, 2019.

\bibitem{romer2019one}
A.~Romer, J.~Berberich, J.~K{\"o}hler, and F.~Allg{\"o}wer, ``One-shot
  verification of dissipativity properties from input--output data,'' {\em IEEE
  Control Systems Letters}, vol.~3, no.~3, pp.~709--714, 2019.

\bibitem{fujisaki2004system}
Y.~Fujisaki, Y.~Duan, and M.~Ikeda, ``System representation and optimal control
  in input-output data space,'' in {\em Proc. of the 10\tss{th} IFAC Symposium
  on Large Scale Systems}, pp.~185--190, 2004.

\bibitem{ikeda2001model}
M.~Ikeda, Y.~Fujisaki, and N.~Hayashi, ``A model-less algorithm for tracking
  control based on input-output data,'' {\em Nonlinear Analysis, Theory,
  Methods and Applications}, vol.~47, no.~3, pp.~1953--1960, 2001.

\bibitem{markovsky2005data}
I.~Markovsky, J.~C. Willems, P.~Rapisarda, and B.~L.~M. De~Moor, ``Data driven
  simulation with applications to system identification,'' in {\em Proc. of the
  16\tss{th} IFAC World Congress}, pp.~970--975, 2005.

\bibitem{markovsky2022interpolation}
I.~Markovsky and F.~D{\"o}rfler, ``Data-driven dynamic interpolation and
  approximation,'' {\em Automatica}, vol.~135, p.~110008, 2022.

\bibitem{markovsky2022data}
I.~Markovsky, ``Data-driven simulation of generalized bilinear systems via
  linear time-invariant embedding,'' {\em IEEE Transactions on Automatic
  Control}, vol.~68, no.~2, pp.~1101--1106, 2022.

\bibitem{mishra2021data}
V.~K. Mishra, I.~Markovsky, A.~Fazzi, and P.~Dreesen, ``Data-driven simulation
  for {NARX} systems,'' in {\em Proc. of the 29\tss{th} European Signal
  Processing Conference}, pp.~1055--1059, 2021.

\bibitem{fazzi2024simulation}
A.~Fazzi and A.~Chiuso, ``Simulation of nonlinear systems trajectories: between
  models and behaviors,'' in {\em Proc. of the 10\tss{th} International
  Conference on Control, Decision and Information Technologies},
  pp.~2049--2054, 2024.

\bibitem{verhoek2024encyclo}
R.~T{\'o}th and C.~Verhoek, ``Modeling and control of {LPV} systems,'' in {\em
  Encyclopedia of Systems and Control Engineering} (Z.~Ding, ed.), vol.~1,
  pp.~405--418, Elsevier, 2026.

\bibitem{verhoek2024flsimple}
C.~Verhoek, I.~Markovsky, S.~Haesaert, and R.~T{\'o}th, ``A behavioral approach
  for {LPV} data-driven representations,'' {\em IEEE Transactions on Automatic
  Control}, pp.~1--14, 2025.

\bibitem{verhoek2023dpcjournal}
C.~Verhoek, J.~Berberich, S.~Haesaert, R.~T{\'o}th, and H.~S. Abbas, ``A linear
  parameter-varying approach to data predictive control,'' {\em arXiv preprint
  arXiv:2311.07140}, 2024.

\bibitem{verhoek2023dissipativity}
C.~Verhoek, J.~Berberich, S.~Haesaert, F.~Allg{\"o}wer, and R.~T{\'o}th,
  ``Data-driven dissipativity analysis of linear parameter-varying systems,''
  {\em IEEE Transactions on Automatic Control}, vol.~69, no.~12,
  pp.~8603--8616, 2024.

\bibitem{markovsky2022identifiability}
I.~Markovsky and F.~D{\"o}rfler, ``Identifiability in the behavioral setting,''
  {\em IEEE Transactions on Automatic Control}, vol.~68, no.~3, pp.~1667--1677,
  2022.

\bibitem{tibshirani1996regression}
R.~Tibshirani, ``Regression shrinkage and selection via the {LASSO},'' {\em
  Journal of the Royal Statistical Society Series B: Statistical Methodology},
  vol.~58, no.~1, pp.~267--288, 1996.

\bibitem{hespe2021convergence}
C.~Hespe and H.~Werner, ``Convergence properties of fast quasi-{LPV} model
  predictive control,'' in {\em Proc. of the 60th Conference on Decision and
  Control}, pp.~3869--3874, 2021.

\end{thebibliography}
